# Photocurrent Properties of Freely Suspended Carbon Nanotubes under Uniaxial Strain

S. M. Kaniber,[1] L. Song,[2] J.P. Kotthaus,[2] and A.W. Holleitner[1,2,]*

1) Walter Schottky Institut and Physik Department, Technische Universität München, Am Coulombwall 3, D-85748 Garching, Germany.

2) Fakultät für Physik and Center for NanoScience, Ludwig-Maximilians-Universität, Geschwister-Scholl-Platz 1, D-80539 München, Germany.

The photocurrent properties of freely suspended single-walled carbon nanotubes (CNTs) are investigated as a function of uniaxial strain. We observe that at low strain, the photocurrent signal of the CNTs increases for increasing strain, while for large strain, the signal decreases, respectively. We interpret the non-monotonous behavior by a superposition of the influence of the uniaxial strain on the resistivity of the CNTs and the effects caused by Schottky contacts between the CNTs and the metal contacts.

Corresponding author: holleitner@wsi.tum.de

Carbon nanotubes (CNTs) have attracted considerable attention because of their compelling optoelectronic[1]-[8] and electro-mechanical properties.[9]-[29] For instance, laser-induced excitonic transitions can give rise to a photoconductance of CNTs.[3],[4] A photoconductance can also be bolometrically induced in CNTs,[6],[8] and surface states due to adsorbates can alter the photoconductance of CNTs by laser-excited photodesorption of the molecules.[2] Furthermore, electric fields at the Schottky contacts between CNTs and metal contacts can separate optically excited electron-hole pairs, causing a photocurrent across electrically contacted CNTs.[5]-[7]

At the same time, the electro-mechanical properties of CNTs have been studied both by locally manipulating CNTs with the tip of an atomic force microscope (AFM),[12]-[14] and by applying uniaxial[15]-[18] and torsional[19],[20] strain to the CNTs. Theorists have modeled the electronic behavior of the mechanically deformed CNTs by an enhanced electronic scattering at defects,[13],[21],[22] a structural induced alteration of the CNTs' band gap,[15]-[19],[21],[23]-[25],[29] and by a mechanical induced transition from $sp^2$ to $sp^3$ hybridization of the carbon bonds.[11],[28]

Here, we report on the photocurrent properties of freely suspended CNTs as a function of statically applied uniaxial strain. To this end, we experimentally verify that the photocurrent is generated at Schottky contacts between the freely suspended CNTs and their bracing source and drain electrodes. Then, the strain is induced by applying a voltage to a piezoelectric stack, such that the distance of the source and drain electrodes is increased. We observe a rise of the photocurrent signal of up to ~150 % for uniaxial strain values in the range of 0.3 to 1.2 % and a decrease of the photocurrent signal for strain values extending this magnitude. We explain the non-monotonous behavior by a superposition of the effect caused by Schottky contacts and the influence of uniaxial strain on the CNTs. In particular, the results at large piezovoltages are consistent with reports that the conductance of the CNTs decreases for increasing mechanical deformation.[10]-[13],[25],[28]

Fig. 1(a) shows a schematic side view of the device for applying strain to the suspended CNTs. Two L-shaped fittings are bonded to both sides of a piezoelectric stack, which extends uniaxially if a voltage $V_{PIEZO}$ is applied to it.[18] The fittings support a silicon substrate which features a center gap of about 70 to 80 μm width. The gap is prepared by optical lithography and KOH etching,[30] such that one side of the gap is open [Fig. 1(b)]. Therefore, the gap expansion around the open side almost corresponds to the piezo expansion. The CNTs are grown via electric-field assisted chemical vapor deposition,[31] such that they are mounted on two Au pads. The pads define the source and drain electrodes for the photocurrent measurements, and they are patterned on an insulating $SiO_2$ layer with a thickness of 100 nm. Scanning electron microscope (SEM) images demonstrate that the suspended CNTs are stretched, when $V_{PIEZO}$ is applied to the piezoelectric stack [Fig. 1(c)-(e)]. In particular, ripples at $V_{PIEZO} = 0V$ [e.g. triangles in Fig. 1(c)] are straightened at $V_{PIEZO} > 0V$, and eventually the CNTs are uniaxially strained [see dashed line and black triangle in Fig. 1(c)-(e)].[18] From the SEM images, we can estimate that the particular sample is stretched by about 1.4 % at $V_{PIEZO} = 35$ V.

Then, photocurrent images of the suspended CNTs are acquired by recording the change of the source-drain current $\Delta I^{DC}$ across the CNTs at a finite source-drain bias $V_{SD}$, when the CNTs are illuminated.[5] To this end, the CNTs are illuminated by focusing the light of a mode-locked titanium:sapphire laser with a photon wavelength in the range of $\lambda_{\mathrm{PHOTON}}$ = 700 nm and 900 nm through the objective of a microscope onto the surface of the sample. The typical laser spot diameter is 2 µm, and the typical power density is ~0.5 kW/cm$^2$. Scanning the laser spot across the CNTs [Fig. 1(b)], the local change $\Delta I^{\mathrm{DC}}(\hat{x},\hat{y}) = I_{ON}^{DC}(\hat{x},\hat{y}) - I_{BACKGROUND}^{DC}$ is detected for the laser being 'on'. $I_{BACKGROUND}^{DC}$ describes the direct current (dc) value, when the samples are illuminated far away from the CNTs. In Fig. 2(a), $\Delta I^{\mathrm{DC}}(\hat{x},\hat{y})$ is plotted using a linear false color scale as a function of the coordinates $\hat{x}$ and $\hat{y}$. We observe two oppositely signed resonances in $\Delta I^{\mathrm{DC}}$ close to the contacts; as can be also inferred from the maximum (circle) and miminum signal (dashed circle) in the corresponding line cut along the $\hat{x}$-coordinate [Fig. 2(b)]. The resonances change their sign, when the source and the drain contacts are switched; i.e., the current amplifier for detecting $\Delta I^{\mathrm{DC}}$ is connected first to drain [Fig. 2(a) and (b)] and then to source [Fig. 2(c) and 2(d)]. Hereby, we confirm recent reports that a Schottky contact between CNTs and metal pads can give rise to a photocurrent.[1],[5],[7] In this process, the electron-hole pairs locally created by photoexcitation are separated due to the local built-in electric field at the Schottky contact [Fig. 2(e)], and a maximum (minimum) photocurrent signal can be detected when the region at the reverse-biased (forward-biased) Schottky contact is illuminated.[5] Furthermore, the interband transitions between the van Hove singularities in the CNTs provide the necessary absorption cross-section,[1] and the generated shortcircuit current manifests itself as an offset voltage in the source-drain dependence of $\Delta I_{\mathrm{DC}}$ [data not shown].

Recent reports show that the resistivity of CNTs increases for increasing mechanical deformation.[10]-[13],[25],[28] In turn, one expects that the photocurrent $I_{ON}^{DC}$, which is optically generated at the Schottky contacts [as depicted in Fig 2(a) and (c)], decreases due to the increased global resistivity of strained CNTs. Fig. 3(a) depicts data of such a strain-induced change of photocurrent as a function of the laboratory time. Here, freely suspended CNTs are optically excited at a position of maximum photocurrent [such as at the circle in Fig. 2(a)] and $V_{\mathrm{PIEZO}}$ is increased from 0V to 30 V in six steps of 5 V. We detect that $I_{ON}^{DC}$ decreases monotonously for increasing voltage steps. The measurement is reversible in a way, that $I_{ON}^{DC}$ reaches similar values for each step VI to I, when $V_{\mathrm{PIEZO}}$ is decreased from 30V to 0V [data not shown]. Hereby, we interpret the decrease of $I_{ON}^{DC}$ for increasing $V_{\mathrm{PIEZO}}$ to reflect an increase of the CNTs' global resistivity $\rho_{\mathrm{CNT}}$ due to mechanical deformation.[10]-[13],[25],[28]

To increase the experimental sensitivity of the photocurrent measurements, we chop the exciting laser field at a frequency $f_{\mathrm{CHOP}}$ and additionally amplify the resulting current $\Delta I^{LOCK-IN} = I_{ON}^{LOCK-IN}(\hat{x}, \hat{y}, f_{\mathrm{CHOP}}, V_{\mathrm{PIEZO}}) - I_{OFF}^{LOCK-IN}(f_{\mathrm{CHOP}}, V_{\mathrm{PIEZO}})$ across the sample for the laser being "ON" or "OFF", respectively, by a current-voltage converter in combination with a lock-in amplifier at the reference frequency $f_{\mathrm{CHOP}}$. This technique allows us to distinguish a small optically induced change of $\Delta I^{LOCK-IN}$ as a function of $V_{\mathrm{PIEZO}}$ [~ a few tens of pA per step in Fig. 3(a)] from the larger change of $I_{ON}^{DC}$, which is induced by the global change of the CNTs' resistivity as a function of $V_{\mathrm{PIEZO}}$ [~ 260 pA per step in Fig. 3(a)]. Fig. 3(a) depicts a simultaneous measurement of $\Delta I^{\mathrm{LOCK-IN}}$ and $I_{ON}^{DC}$.

We find that $\Delta I^{\text{LOCK-IN}}$ first increases (step I) and then decreases (steps II to VI). Again, the measurement of $\Delta I^{\text{LOCK-IN}}$ is reversible, when $V_{\text{PIEZO}}$ is decreased from 30V to 0V [data not shown]. For large strain values, we can assume that the increased resistivity equally dominates the amplitudes $I_{ON}^{LOCK-IN}$ and $I_{OFF}^{LOCK-IN}$. In turn, we can explain the decrease and the saturation of $\Delta I^{\text{LOCK-IN}} = I_{ON}^{LOCK-IN} - I_{OFF}^{LOCK-IN}$ for large strain values phenomenologically [steps IV to VI of Fig 3(a)]. From SEM images such as in Fig. 1(c) to (e), we can estimate the strain value for the maximum photocurrent to be ~ 0.3 % at $V_{\text{PIEZO}}$ = 5 V [step I in Fig. 3(a)]. Most strikingly, the relative change of $\Delta I^{\text{LOCK-IN}}$ as a function of $V_{\text{PIEZO}}$ is significantly larger than the one of $I_{ON}^{DC}$; i.e. $\Delta I^{\text{LOCK-IN}}(V_{\text{PIEZO}} = 5\text{ V}) / \Delta I^{\text{LOCK-IN}}(V_{\text{PIEZO}} = 0\text{ V}) \approx 148\ \%$, $\Delta I^{\text{LOCK-IN}}(V_{\text{PIEZO}} = 30\text{ V}) / \Delta I^{\text{LOCK-IN}}(V_{\text{PIEZO}} = 0\text{ V}) \approx 55\ \%$, and $I_{ON}^{DC}(V_{\text{PIEZO}} = 30\text{ V}) / I_{ON}^{DC}(V_{\text{PIEZO}} = 0\text{ V}) \approx 94\ \%$. This observation makes it plausible that the origin of the photocurrent, i.e., the built-in electric fields close to a Schottky contact, is altered when the samples are stretched.

To substantiate the last interpretation we detect photocurrent images of $\Delta I^{\text{LOCK-IN}} = \Delta I^{\text{LOCK-IN}}(\hat{x}, \hat{y})$ of a bundle of CNTs close to the location of maximum photocurrent at different $V_{\text{PIEZO}}$. For the particular bundle of CNTs characterized in Fig. 3(b) and (c), the region of maximum photocurrent at $V_{\text{PIEZO}}$ = 0 V is extended towards the ‘left’ [Fig. 3(b)]. The observation of an extended ‘lobe structure’ is consistent with the interpretation, that in a CNT bundle the built-in electric fields due to a Schottky contact are extended to a large region because of intertube transitions with different morphology and varying space charge configurations.[5] Most significantly, at $V_{\text{PIEZO}}$ = 20 V corresponding to an estimated strain value ~1.2 %, $\Delta I^{\text{LOCK-IN}}$ exhibits locally a new maximum [circle in Fig. 3(c)], which is larger than the photocurrent signal of the general 'lobe structure'. We interpret the new maximum at $V_{\text{PIEZO}} > 0$V to reflect large electric fields at the location of maximum strain within the CNT bundle. In Fig. 3(d), the maximum value of $\Delta I^{\text{LOCK-IN}}$ within the circle in Fig. 3(b) and (c) is plotted as a function of $V_{\text{PIEZO}}$. We detect that the local maximum value of $\Delta I^{\text{LOCK-IN}}$ first rises and then decreases for increasing $V_{\text{PIEZO}}$, which is consistent with the observations in Fig. 3(a). Hereby, we interpret the non-monotonous behavior resulting from a superposition of the local effect caused by a mechanically deformed contact region forming a Schottky contact and the globally increasing effect of uniaxial strain on the resistivity of the CNTs.

We would like to note that the increase of $\rho_{\text{CNT}}$ as a function of the uniaxial strain is usually explained by a structural induced alteration of the CNTs' band gap.[15]-[19],[23]-[25],[29] Recent measurements verified this band gap change to be in the range of about 12 meV for a maximum strain value and a device geometry similar to the one reported here.[18] We have repeated the photocurrent measurements as a function of the uniaxial strain at different laser wavelengths $\lambda_{\text{LASER}}$ [data not shown]. First, we detect a spectrally broad maximum of the photocurrent at $\lambda_{\text{LASER}} = 830 \pm 20$ nm at $V_{\text{PIEZO}} = 0$ V, which is consistent with a CNT diameter of about $d_{\text{CNT}} = 1.10 \pm 0.05$ nm.[8] Second, we do not detect any strain induced shift of the resonant excitation energy, because the broad photocurrent maximum at $V_{\text{PIEZO}} = 0$ V seems to mask a possible shift of the photocurrent resonance at $V_{\text{PIEZO}} > 0$ V.

In summary, we present spatially resolved photocurrent measurements on freely suspended CNTs. The CNTs are strained by elongating a piezoelectric stack attached to the sample. We observe a rise of the photocurrent signal of up to ~150% for uniaxial strain of about 0.3 - 1.2 % and a decrease of the photocurrent signal for strain values exceeding this value. We explain the non-monotonous behavior by a superposition of a local effect caused by Schottky contacts and the global influence of uniaxial strain on the CNTs.

We gratefully acknowledge financial support by the DFG (SFB 486 TPA1 and Ho 3324/4), the Center for NanoScience (CeNS), the LMUexcellent program and the German excellence initiative via the "Nanosystems Initiative Munich" (NIM). L.S. thanks the Alexander von Humboldt foundation for their support.

**Fig. 1** Schematic side **(a)** and top **(b)** view of a $Si/SiO_2$ sample with a "T"-slit mounted on a piezoelectric stack. Freely suspended single-walled carbon nanotubes (CNTs) bridge the gap between the source and drain electrodes. Applying a voltage $V_{PIEZO}$ to the piezoelectric stack allows applying uniaxial strain to the CNTs. **(c)** to **(e)**: Scanning electron microscope (SEM) images of the CNTs at $V_{PIEZO}$ = 0 V, 15 V, and 30 V at room temperature (see text for details).

**Fig. 2(a)** Photocurrent image of freely suspended CNTs; i.e. change of the source-drain current $\Delta I_{DC}$ when the CNTs are illuminated as a function of the laser position ($\lambda_{LASER}$ = 830 nm, $V_{SD}$ = -50 mV, room temperature). A minimum (dashed circle) is located close to the left metal contact, while close to the right metal contact a maximum occurs (circle). **(b)** Single trace along the dashed line in Fig. 2(a). **(c)** Photocurrent image for the reversed bias as in (a) ($\lambda_{LASER}$ = 800 nm, $V_{SD}$ = +50 mV, room temperature). **(d)** Single trace along the dashed line in Fig. 2(c). **(e)** Schematic of Schottky barriers between the CNTs and the contacting metal electrodes.

**Fig. 3 (a)** Photocurrent $\Delta I^{\text{LOCK-IN}}$ and time integrated current $I_{ON}^{DC}$ across freely suspended CNTs as a function of laboratory time, when $V_{PIEZO}$ is increased from 0 V to 30 V in six steps of 5 V ($\lambda_{LASER}$ = 714 nm, $V_{SD}$ = +50 mV, $f_{CHOP}$ = 912 Hz, room temperature). **(b)** and **(c)** Photocurrent image of $\Delta I^{\text{LOCK-IN}}$ of a bundle of CNTs close to a metal contact ($\lambda_{LASER}$ = 714 nm, $V_{SD}$ = +50 mV, $f_{CHOP}$ = 912 Hz, room temperature). **(d)** Maximum values of $\Delta I^{\text{LOCK-IN}}$ taken from the encircled area in Fig. 3(b) and (c) and further photocurrent scans as a function of $V_{PIEZO}$.

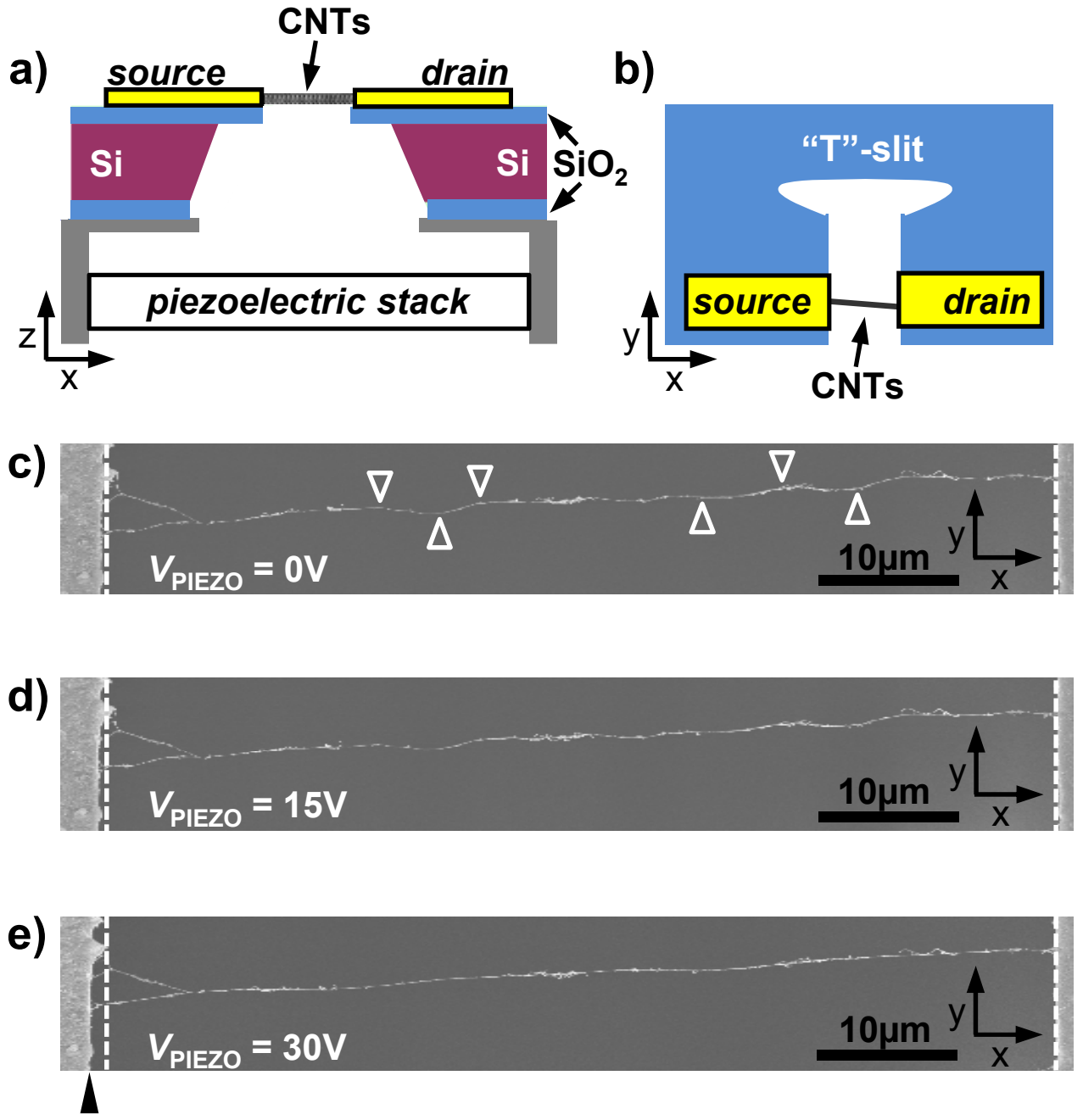

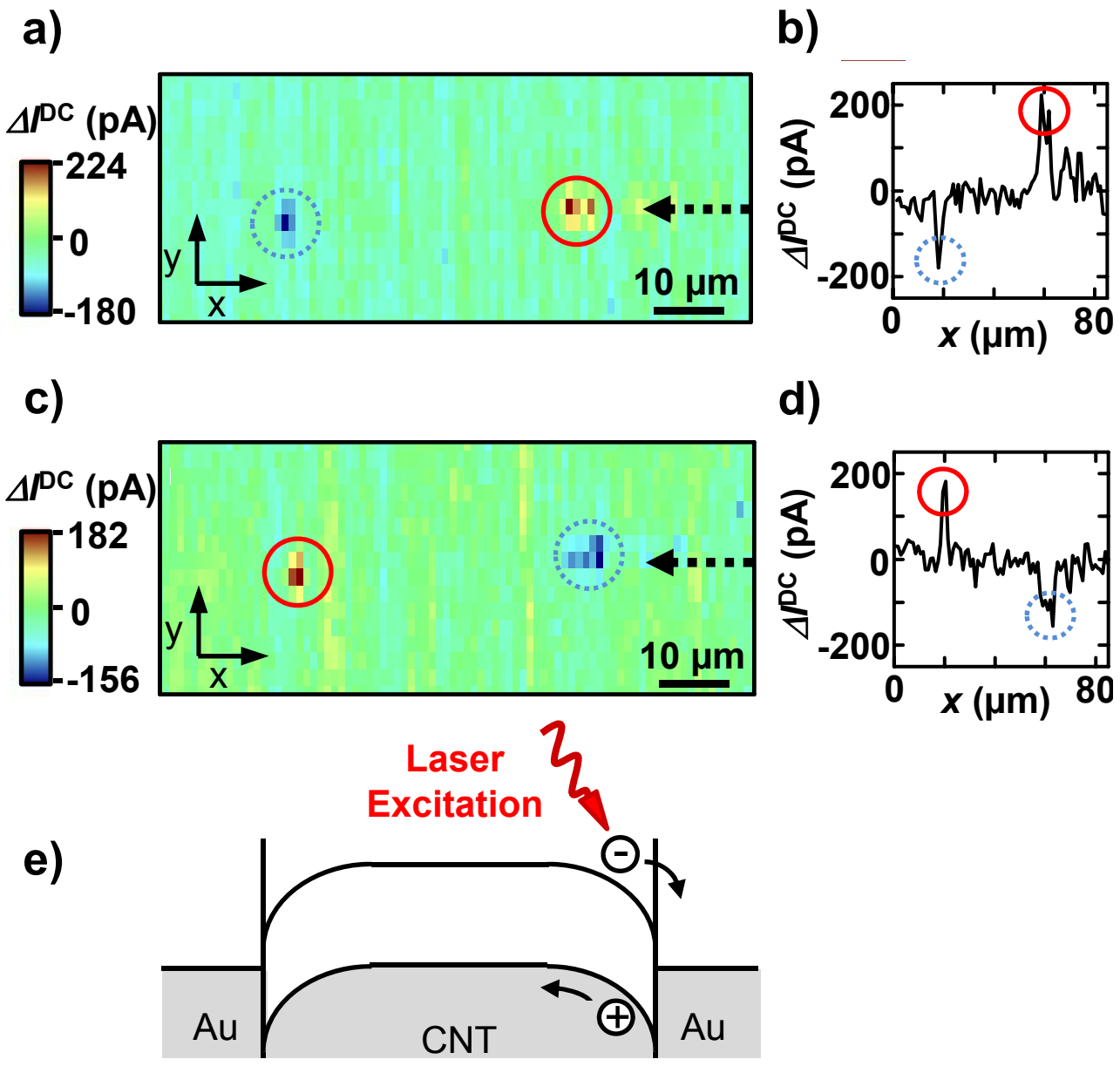

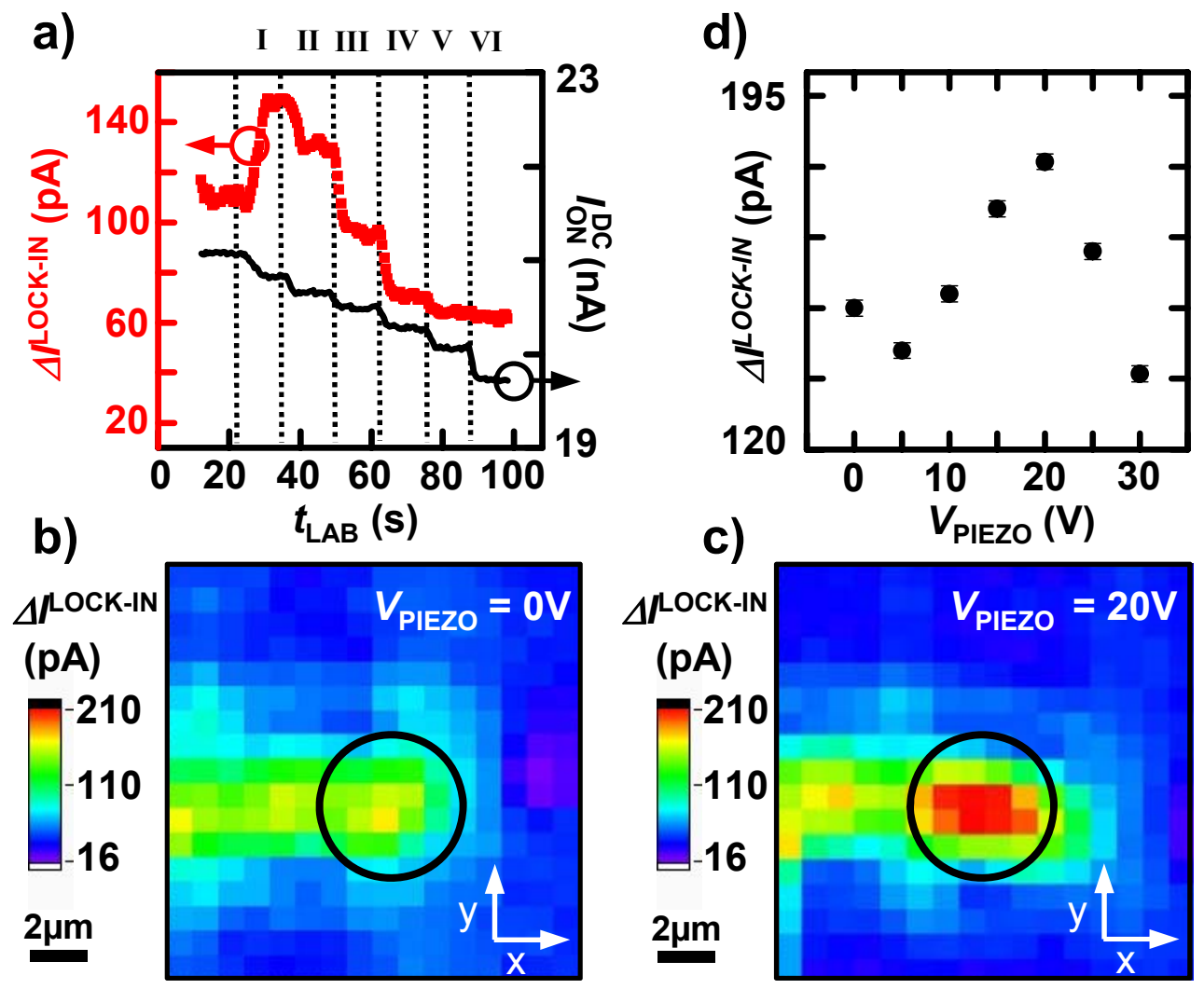